# Degradation of layered oxide cathode in a sodium battery: a detailed investigation by X-ray tomography at the nanoscale

Daniele Di Lecce,[a,b,c] Vittorio Marangon,[d] Mark Isaacs,[e,f] Robert Palgrave,[e,f] Paul R. Shearing,[b,c]
Jusef Hassoun[a,d,g]*

[a] *Graphene Labs, Istituto Italiano di Tecnologia, via Morego 30, Genova, 16163, Italy*

[b] *Electrochemical Innovation Lab, Department of Chemical Engineering, UCL, London, WC1E 7JE, United Kingdom*

[c] *The Faraday Institution, Quad One, Becquerel Ave, Harwell Campus, Didcot, OX11 0RA United Kingdom*

[d] *University of Ferrara, Department of Chemical, Pharmaceutical, and Agricultural Sciences, Via Fossato di Mortara 17, Ferrara, 44121, Italy*

[e] *Department of Chemistry, UCL, 20 Gordon St, Bloomsbury, London, WC1H 0AJ, United Kingdom*

[f] *HarwellXPS, Research Complex at Harwell, Rutherford Appleton Laboratories, Harwell, Didcot, OX11 0FA, United Kingdom*

[g] *National Interuniversity Consortium of Materials Science and Technology (INSTM), University of Ferrara Research Unit, University of Ferrara, Via Fossato di Mortara, 17, 44121, Ferrara, Italy.*

* Corresponding author: jusef.hassoun@unife.it, jusef.hassoun@iit.it.

## Abstract

The degradation mechanism in sodium cell of a layered $Na_{0.48}Al_{0.03}Co_{0.18}Ni_{0.18}Mn_{0.47}O_2$ cathode with P3/P2 structure is investigated by revealing the changes in microstructure and composition upon cycling. The work aims to rationalize the gradual performance decay and the alteration of the electrochemical response in terms of polarization, voltage signature, and capacity loss. Spatial reconstructions of the electrode by X-ray computed tomography at the nanoscale supported by quantitative and qualitative analyses show fractures and deformations in the cycled layered metal-oxide particles, as well as inorganic side compounds deposited on the material. These irreversible morphological modifications reflect structural heterogeneities across the cathode particles due to formation of various domains with different Na$^+$ intercalation degree. Besides, X-ray photoelectron spectroscopy data suggest that the latter inorganic species in the cycled electrode are mainly composed of NaF, Na$_2$O, and NaCO$_3$ formed by parasitic electrolyte decomposition. The precipitation of these insulating compounds at the electrode/electrolyte interphase and the related structural stresses induced in the material lead to a decrease in cathode particle size and partial loss of electrochemical activity. The retention of the Na$_{0.48}$Al$_{0.03}$Co$_{0.18}$Ni$_{0.18}$Mn$_{0.47}$O$_2$ phase after cycling suggests that electrolyte upgrade may improve the performance of the cathode to achieve practical application for sustainable energy storage.

**Table of Content**

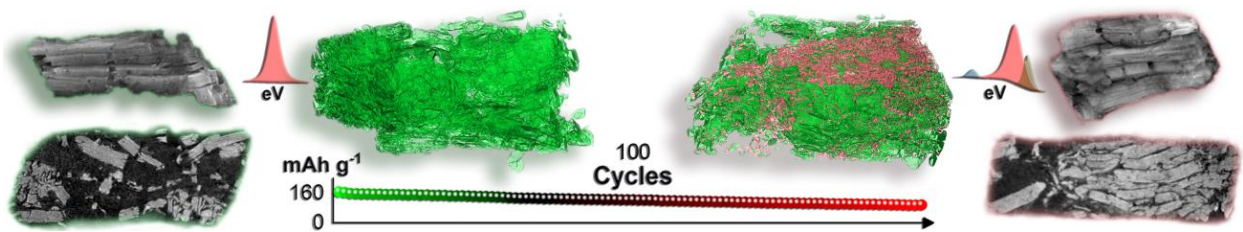

**A sodium layered oxide cathode** is observed with relevant detail using high-resolution X-ray computed tomography to disclose the reasons of capacity fade upon long-term cycling in the battery, and additional findings are given by X-ray photoelectron spectroscopy.

## Introduction

Sodium-ion batteries are currently gaining a great deal of attention as viable alternatives to conventional lithium-ion systems,[1,2] in view of outstanding results recently obtained at the laboratory scale.[3,4] The investigation of sodium batteries began alongside pioneering studies demonstrating the electrochemical intercalation of alkali ions in chalcogenides,[5] which were driven by the high theoretical energy density of lithium, sodium, and potassium, having a redox potential *vs* standard hydrogen electrode (SHE) of −3.04 V, −2.71 V, and −2.93 V, respectively, as well as a relatively low weight.[6] The concept of the "rocking-chair battery" introduced in 1980[7] was supported by groundbreaking research on intercalation electrodes[5] and aprotic electrolyte media[8] which led to the commercialization of the first lithium-ion battery in 1991.[9] This technology has become crucial in our society, being exploited in a vast variety of portable electronics, including smartphones, laptops and tablets, as well as in electric vehicles (EVs) and in stationary energy storage systems. Therefore, the Li-ion battery has gained increasing interest in the scientific community over recent years,[10] as acknowledged by the Nobel Prize in Chemistry 2019 rewarded to M. S. Whittingham, J. B. Goodenough, and A. Yoshino.[11] On the other hand, the widespread diffusion of this intriguing electrochemical system and the growing demand from emerging markets, such as the electric and hybrid vehicles and the renewable energy sector, are presently raising concerns about possible unbalances between consumption and production of raw materials, including high-quality lithium, which might affect their availability and the final cost to the consumer.[12] In principle, the sodium-ion cell involves insertion electrodes that are analogous to those of lithium-ion systems,[13] so this new technology may be developed and manufactured using current battery assembly lines.[1] Sodium is widely abundant, has a relatively low cost as compared to lithium, and allows the use of aluminum instead of copper as a current collector at the negative side.[13] Encouraging results have demonstrated that sodium-ion intercalation systems using thin-film electrodes and aprotic electrolyte solutions may exhibit very promising electrochemical behavior;[14,15] however, a lower energy density than that of the lithium-ion counterpart is still an open issue.[16] In this scenario, the development of sodium-ion batteries may actually mitigate the price per kWh of electrochemical energy storage, which is particularly relevant to large-scale stationary applications.[17] Among the various cathode materials for sodium batteries investigated so far,[13] layered metal oxides with crystal structures similar to that of conventional $LiCoO_2$ and Ni-rich derivatives, e.g., $LiNi_{0.8}Mn_{0.1}Co_{0.1}O_2$ (NMC 811) and $LiNi_{0.8}Co_{0.15}Al_{0.05}O_2$ (NCA), have shown high performance, particularly in terms of delivered capacity, both in half and in full cells.[3] These compounds may have the general formula $Na_{1-x}MO_2$, where M is a transition metal, such as, Mn, Fe, Co, Ni, Ti, V, or Cr and $x$ may vary from low values to unity.[3] Despite being capable of reversibly reacting in sodium cells with suitably fast kinetics and specific capacity, these layered cathodes may suffer from insufficient structural stability upon repeated charge/discharge cycles, mostly due to phase transitions promoted by the (de)intercalation of $Na^+$ ions.[18] Typically, structural changes in the positive electrode during the electrochemical process can stepwise alter the voltage profiles and hinder the long-term cycling ability of the cell, whilst additional activation energy for phase boundary movement generally limits the rate capability.[18,19] In this regard, careful tuning of the $Na_{1-x}MO_2$ stoichiometry *via* incorporation of various transition metals[20,21] and Al-doping[22] can actually stabilize the crystal lattice and enhance the high-voltage electrochemical processes. Accordingly, we have recently optimized a transition metal-substituted $Na_{0.48}Al_{0.03}Co_{0.18}Ni_{0.18}Mn_{0.47}O_2$ (NCAM) layered cathode reversibly reacting in sodium cells by a single-phase, solid-solution mechanism with smooth potential curve and rather constant slope within the wide range from 1.4

V *vs* Na$^+$/Na to 4.6 V *vs* Na$^+$/Na, delivering a reversible capacity of about 175 mAh g$^{-1}$.[23] Such a structurally optimized multi-metal formulation ensured a very promising performance and high reversibility upon the initial charge/discharge cycles, avoiding phase transitions upon Na$^+$ (de)intercalation and mitigating the Jahn-Teller distortion on Mn$^{3+}$.[23] Furthermore, the inclusion of Co and Ni may improve the electrode operation in the cell above 3 V *vs* Na$^+$/Na,[21] whilst aluminum has proven to stabilize the structure at high voltage, despite being electrochemically inactive.[22] However, the capacity decay over cycling was not fully mitigated.[23] Therefore, we aim in this work to investigate fully the degradation mechanisms of NCAM during cycling in the sodium cell by a comprehensive approach principally based on the three-dimensional (3D) reconstructions of the cathode before and after cycling by X-ray computed tomography (CT). This alternative approach enables us to display the evolution of the NCAM particles at the nanoscale and the changes in electrode microstructure. Indeed, X-ray CT provides qualitative and quantitative data on the spatial distribution of the various electrode components, which are identified by local differences in attenuation of the X-ray beam,[24–26] thereby allowing a direct observation of the defects formed within the layered NCAM domains, along with possible changes in the particle size after cycling. The analysis is also supported by X-ray photoelectron spectroscopy (XPS) and electron microscopy data to detect the morphology and composition of the electrode/electrolyte interphase (EEI). Hence, our study sheds further light on various phenomena driving the performance of layered oxide cathodes and suggests potential strategies to achieve sodium-ion batteries with enhanced cycle life and practical interest.

**Results and discussion**

Sodium cells employing the NCAM cathode may undergo a gradual deterioration during cycling, which may be reflected as capacity decay along with alteration of the characteristic voltage profile.[23] Accordingly, the Na/NCAM cell reported in Figure 1a evidences a decrease in reversible capacity from 170 mAh g$^{-1}$ to 100 mAh g$^{-1}$ after 100 cycles, that is, about a 40% loss, whilst the coulombic efficiency values range from 97.5% to 99.6% (the average efficiency is 98.4%), thus suggesting a parasitic electrolyte decomposition.[2] Typically, irreversible processes take place across the whole cell upon cycling, and affect the stability of the interphases on both anode and cathode.[2] In this regard, it is well known that the highly reactive sodium-metal anode may cause reduction of the electrolyte species with deposition of a thick solid electrolyte interphase (SEI) layer hindering the electrode charge transfer.[27] Moreover, electrolyte oxidation may occur over the layered oxide cathode at high electrochemical potential *vs* Na$^+$/Na[28] along with possible irreversible phase transitions inhibiting a smooth Na$^+$ intercalation process in the electrode bulk.[13] However, we have demonstrated in an earlier work that the NCAM material reacts in sodium cells according to a single-phase process, benefiting from an optimized stoichiometry and a stable P3/P2 structure, which lead to a characteristic profile with rather constant slope as well as to high reversibility upon a limited number of charge/discharge cycles.[23] The new test prolonged herein over 100 cycles (Figure 1a) reveals a capacity loss ascribable only in part to undesired reactions taking place at the anode side. Indeed, the cell shows a significant modification of the voltage signature and increase of the polarization possibly associated with microstructural reorganizations or gradual insulation of the NCAM material (Figure 1b). This change in voltage signature is clearly evidenced by the differential capacity plots (Figure 1c) which show a remarkable shift with broadening of the peaks after 50 and 100 cycles, and likely indicate an altered multistep intercalation process involving the Mn$^{4+}$/Mn$^{3+}$, Ni$^{4+}$/Ni$^{2+}$, and Co$^{4+}$/Co$^{3+}$ redox couples at increasing potential *vs* Na$^+$/Na.[23]

**Figure 1**

The detrimental effects on the battery behavior of possible parasitic reactions between sodium metal and the electrolyte solution during cycling have been extensively described in the literature.[2,27,29] Therefore, the investigation of the negative electrode will be only partially considered in this work, in particular taking into account a possible replacement of the sodium metal with a more suitable Na-ion anode to mitigate these adverse processes.[4,30–34] Instead, we have principally focused our attention on unravelling microstructural changes in the NCAM cathode that may further jeopardize the performance of the cell. Substantial insight on the evolution of the positive electrode is gathered herein by combining quantitative and qualitative data of X-ray diffraction (XRD), electron microscopy, and XPS analyses with relevant results of X-ray CT investigations at the nanoscale. Figure 2 shows scanning electron microscopy (SEM) images (a, b) and XRD patterns (c-f) of the electrode film before and after cycling. The pristine NCAM clearly reveals the layered morphology of the oxide particles which are embedded in a carbon-binder matrix of the positive electrode film (Figure 2a). Relevantly, the layered-oxide morphology is somewhat preserved after 100 cycles that indicates partial microstructural retention, although cracks and irregularities, as well as slight bending of the sheets, are barely detectable (Figure 2b). Hence, orderly stacked oxide layers in the pristine NCAM (Figure 2a) arrange into several domains with different sheet orientation in the cycled NCAM (Figure 2b), thus suggesting locally inhomogeneous Na$^+$ intercalation which may cause structural stresses and possible crakes during cycling.[35,36] SEM images of the electrode after 50 cycles reveal a similar layered morphology (see Figure S1 in the Supporting Information).

Rietveld refinement of XRD data has suggested a P3/P2 volume ratio of 79:21.[23] On the other hand, the diffraction patterns of the electrode film (Figure 2c–f) do not clearly show the minor reflections of the secondary phase, that is, the P2 structure, and have a higher signal-to-noise ratio as compared to that obtained from the powder spread on a sample holder.[23] In this regard, it is worth mentioning that various sample characteristics, *e.g.*, amount of powder, presence of electrode additives, sample thickness, may influence the quality of the diffraction data.[37] The XRD patterns of the electrode before and after cycling (Figure 2c–f) evidence multiple changes promoted by the electrochemical process, among which the most relevant are represented by: *i)* the splitting of the (003) and (002) reflections associated with the P3 and P2 phases, respectively (Figure 2d); *ii)* a shift towards higher angles of the (006) and (004) reflections and concomitantly the appearance of two additional broad peaks at 32.5° and 32.8° (Figure 2e), *iii)* a shift towards lower angles of the (101), (012), and (015) reflections of the major P3 phase (Figure 3f). In detail, a shift to high $2\theta$ of the (00$n$) reflections indicates a contraction of the $c$ parameter of the P3 and P2 unit cells, that is, a decrease in distance between the slabs formed by edge-sharing $MO_6$ octahedra (where M is a transition metal atom).[23,38,39] This phenomenon typically occurs due to shielding of the negatively charged slabs upon $Na^+$ intercalation in the oxide framework, along with an increase of the $a$ parameter ascribed to a lower oxidation state of the transition metal center causing an increase in ionic radius.[23,38,39] The detailed analysis of the XRD patterns reported in Figure S2 in the Supporting Information suggests the absence of structural rearrangements deriving from phase transitions during the electrochemical process, although possible traces of P1 domains perhaps formed at low sodium content cannot be fully excluded.[40] On the other hand, XRD measurements have already shown that NCAM reacts in sodium cells via a solid-solution mechanism within the wide potential range from 1.4 V *vs* $Na^+$/Na to 4.6 V *vs* $Na^+$/Na.[23] Therefore, the above-mentioned peak splitting and occurrence of additional reflections in the XRD pattern can be reasonably attributed to the existence of several P3/P2 domains having different $Na^+$ intercalation degree in the oxide cathode,[41] rather than to nucleation of new phases during cell operation. In spite of the above local heterogeneities, the overall decrease in interlayer distance and expansion along the plane of the $MO_6$ slabs observed after cycling are in full agreement with the change in average intercalation degree of $Na_xAl_{0.03}Co_{0.18}Ni_{0.18}Mn_{0.47}O_2$, from $x = 0.48$ in pristine condition to $x \approx 0.80$ at the end of the discharge step at 1.4 V exploited herein.[23] Given an optimal ion transport into the electrolyte solution,[42,43] a limited sodiation uniformity in the positive electrode[41] and possible deposition of undesired compounds may still affect the NCAM framework conductivity and hinder $Na^+$ and electron transport.[28] Accordingly, the energy dispersive spectroscopy (EDS) maps of the NCAM before and after cycling reported in Figure S3 in the Supporting Information evidence the presence of micrometric agglomerates of sodium-containing species over the electrode surface, and elemental quantification (Table 1) reveals a large increase in F and Na content, as well as depletion of C.

**Figure 2**

**Table 1**

X-ray CT at the nanoscale corroborates the hypotheses drawn by XRD and SEM-EDS analyses, and reveals that cycling significantly affects the electrode microstructure. Although XRD provides unambiguous evidence of structural changes in the cathode, SEM may only reveal the surface morphology of the electrode, which gives us partial information on its microstructural evolution during cycling. On the other hand, X-ray CT enables a detailed reconstruction of the whole cathode sample at the nanoscale. Figure 3 reports cross-sectional slices extracted in a plane parallel to the specimen rotation axis for the electrode before (panel a) and after 100 cycles (panels c), with the corresponding image segmentation (panels b and d, respectively). The local attenuation of the X-ray beam, visualized using a grayscale in Figure 3a and c, provides information on the density of the various phases forming the cathode film.[24–26] This insight enables us to unambiguously discern the NCAM particles (light gray in Figure 3a and c, and green in Figure 3b and d) which are embedded in the carbon-binder mixture (dark gray in Figure 3; the current collector has been removed upon sample preparation as described in the experimental section). The 3D tomographic reconstruction reveals the characteristic microstructural features of the pristine electrode bulk which consists of NCAM sheets stacked into particles with various size (Figure 3a). In this regard, Supporting Movie S1 shows the cross-sectional slices parallel to that displayed in Figure 3a (left side) and corresponding 3-phase segmented stack (right side). The above movie also shows the presence of some delamination across the NCAM domains within the entire volume which might increase the material porosity, enhance the electrode/electrolyte contact, and expose the particle interior (to the electrolyte) thereby possibly improving the charge transfer kinetics.[23,44] On the other hand, the electrode after 100 cycles is characterized by large agglomerates of NCAM particles with appear more compact than those of the pristine material, as clearly shown in Figure 3c as well as in Supporting Movie S2. Remarkably, such an apparent such an apparent increase of compactness degree may be associated with the structural changes detected by XRD and described above.[23,38,39] Indeed, the pristine NCAM particles appear largely flat and mainly composed of regularly packed metal oxide slabs (Figure 3a and Supporting Movie S1), whilst the cycled NCAM incorporate fragments with different sheet orientation, bent and apparently warped in specific regions (Figure 3c and Supporting Movie S2). According to these results, prolonged cycling in the sodium cell may lead to significant deformation of the metal oxide particles and morphology change, likely due to a different $Na^+$ intercalation degree across the material, which is in full agreement with the results of the XRD and SEM analyses of Figure 2.[41]

Besides, X-ray CT at the nanoscale reveals in the electrode after 100 cycles the appearance of an additional phase with lower density than that of NCAM, without the typical morphology ascribed to the layered transition metal oxide cathode. This phase is represented by intermediate grayscale values in Figure 3c and Supporting Movie S2 (left side), since it exhibits moderate X-ray attenuation as compared to NCAM (highly attenuating) and carbon-binder (lowly attenuating). Indeed, this additional domain is likely promoted by cycling and highlighted in light red color in the segmented slice of Figure 3d and Supporting Movie S2 (right side), whilst the related attenuation histograms of the X-ray CT datasets are shown in Figure S4 of the Supporting Information. Such high-resolution datasets have enabled us to reconstruct in detail volume renderings of the NCAM particles along with the additional phase, as shown in Figure 3e and f. This image reveals that after 100 cycles the active material (green phase in panels e and f) is partially covered by the above discussed domain (light red phase in panel f), which is distributed across the whole electrode volume. Further visualization of the cathode reconstruction before and after cycling is provided in Supporting Movies S3 and S4, respectively. Taking into account the relative density as well as the results of EDS analyses (Table 1), we propose that the additional phase might be attributable to inorganic compounds formed upon cycling by side reaction of the NCAM with the propylene carbonate-fluoroethylene carbonate-$NaClO_4$ (PC-FEC-$NaClO_4$) solution at the electrode/electrolyte interphase, subsequently indicated by the acronym EEI in Figure 3 and Supporting Movie S4.[28] We would remark that the NCAM/electrolyte interphase is polarized herein up to an electrochemical potential as high as 4.6 V $vs$ $Na^+$/Na in order to achieve a suitable energy density (see Figure 1b–c), which is a rather high potential (i.e., 4.9 V) if referred to $Li^+$/Li for a reasonable comparison. This relatively high charge limit may actually pose additional challenges in regard to the electrochemical stability of the electrolyte solution.[28] Thus, minor electrolyte decomposition over 100 cycles might cause precipitation of various compounds on the cathode surface, such as NaF, $Na_2O$ and $Na_2CO_3$,[28] which affect the coulombic efficiency values of Figure 1a. This hypothesis is indeed partly supported by the significant increase in F and Na content suggested by EDS (Table 1). Furthermore, the precipitation of these inorganic species over the electrode surface may mask the signal attributed to binder and carbon black additive, thus decreasing the amount of C observed by EDS after cycling (Table 1). On the other hand, the volume rendering of Figure 3f and the reconstruction in Supporting Movie S4 suggest that the EEI species are heterogeneously distributed over the NCAM phase, thereby partially insulating the active material domains, locally affecting the $Na^+$ exchange kinetics, and finally leading to the gradual capacity decay by cycles observed in Figure 1a as well as the voltage profile modification displayed in Figure 1b.[28,41] X-ray imaging of the NCAM electrode after 50 cycles fully confirms the trend of microstructural evolution discussed above (see Figure S5 in the Supporting Information).

**Figure 3**

Quantitative analyses of the X-ray CT datasets further clarify the microstructural modifications occurring in the cathode and reveal a change in distribution of NCAM particles after cycling. Figure 4 shows the "continuous particle size distribution" (PSD) of the layered metal oxide before and after cycling (panels a and b, respectively), where the segmented binary data are processed to estimate the amount of spheres with radius $r$ ($x$ axis) that might fill the volume of analyzed phase.[45] The method to calculate the PSD used in this work provides results that may differ significantly from those obtained by conventional "discrete PSD" approaches.[45] Major deviations are indeed observed between the microstructural data on agglomerates of particles gathered by these different PSD analyses, as the latter distribution only considers the volume of each isolated particle, aggregate, and cluster rather than shape (e.g., elongated or approximately spherical particles) and agglomeration degree (e.g., large, isolated particles or small, interconnected particles). Notably, the "continuous PSD" may more accurately describe agglomerates of small particles, thereby drastically mitigating the effects of instrumental resolution and segmentation artifacts on the results of image-based analyses.

Figure 4 evidences a decrease in average domain size after repeated $Na^+$ (de)insertion in the oxide lattice, and shows that the distribution of cycled NCAM is narrower than that of the pristine compound (compare panels a and b). According to this PSD analysis, 80% of NCAM domains have an estimated radius below 1.3 μm in pristine condition and below 1.0 μm after the cycling test. Notably, such a decrease in average size of NCAM domains after prolonged charge/discharge cycles is in full agreement with the formation of cracks in the oxide particles observed by SEM (Figure 2a and b), and the partial aggregation of various adjacent particles having different orientation observed by X-ray CT imaging (Figure 3a and c). Similar results are obtained by the analysis of the electrode after 50 cycles (Figure S6 in the Supporting Information). Therefore, our data suggest that cell operation induces deformations of the metal oxide layers with consequent particle cracking which are principally promoted by the heterogeneity of $Na^+$ intercalation degree across the material. Apparently, the EEI species formed upon cycling play a key role in steering these microstructural modifications by locally hindering a smooth electrode charge transfer, thus partially limiting the homogeneity of the $Na^+$ intercalation degree.[28] As expected, this EEI domain mainly includes particles smaller than 400 nm, as suggested by a quantitative analysis of the X-ray CT data provided in the Supporting Information (PSD in Figure S7).

**Figure 4**

XPS analyses of pristine and cycled NCAM electrode samples may further elucidate the nature of the EEI species. In this regard, the survey spectra of Figure 5a reveal that both specimens have a complex chemistry and indicate a notable change in surface composition after cycling.

Indeed, the spectrum of pristine NCAM shows the signatures of the main elements forming the oxide, i.e., Na,[46] Co,[47] Mn, Ni,[48] and O,[49] as well as those of F and C attributed to polymer binder (polyvinylidene fluoride, PVDF) and conductive agent (carbon black),[50,51] whilst the Al 2p doublet typically occurring at about a binding energy (BE) of 73 eV is marginally detected and evidenced in the selected region of Figure S8a of the Supporting Information.[52] The spectrum after cycling (Figure 5a) evidences a substantial raise of the sodium peaks along with a notable attenuation of the transition metal peaks, which may be causally related to formation of precipitates over the electrode surface as above described. Accordingly, elemental quantification indicates a massive increase in the amount of Na, moderate raise in O and F content, and depletion of C (Table 1). We remark that these data reasonably agree with the EDS quantification in suggesting deposition of an EEI incorporating sodium, fluorine, and oxygen, despite the intrinsic difference of the two techniques in terms of experimental setup and sample characteristics. In this regard, possible side changes in surface composition due to exposure of the specimens to the atmosphere before the SEM-EDS analyses cannot be excluded (see the experimental section for further details).

The selected regions of the Na 1s, F 1s, O 1s, and C 1s photoelectron signals shown in the deconvoluted spectra of Figure 5b–i enable to identify the main species forming the EEI. Panels b and c of this figure display a single Na 1s peak at 1071.0 eV, reasonably attributed to NCAM, and a F 1s peak at 687.1 eV reflecting the F–C bond of the PVDF binder, along with a shoulder at a lower BE, i.e., 684.7 eV, which indicates fluorine atoms in an ionic environment.[50,51] Our data suggest the presence of NaF deposits on the pristine cathode likely produced via reaction of residual surface species, such as NaOH and $Na_2CO_3$, with PVDF. Notably, sodium layered metal-oxides typically show high affinity towards moisture and $CO_2$, which may result in the formation of sodium hydroxide, sodium carbonate, and transition-metal oxide impurities upon exposure to the atmosphere.[53] Despite being in partial agreement with earlier reports,[46,54] the high NaF signal in the photoelectron spectrum of Figure 5c might indicate either a high fraction of synthesis residues or a particularly severe water sensitivity of NCAM. Accordingly, we have observed slight gelation of the electrode slurry during the coating process, which might suggest the presence of NaOH on the surface of the NCAM particles.[55] Surface coating with metal-oxide thin layers,[53] along with the choice of a more stable polymer binder,[56] might be an effective strategy to solve this issue. On the other hand, the Na 1s BE has approximately the same value in NaF and NCAM so that both photoelectron signals belong to a single peak in Figure 5.[46,57] Furthermore, we may reasonably suppose that the actual amount of NaF in the pristine cathode is sufficiently low to ensure suitable behavior in the sodium cell, as indeed demonstrated by electrochemical tests (Figure 1) and also supported by the thorough study of structure, morphology, and elemental composition discussed above, as well as by earlier reports on the synthesis of NCAM.[23] According to the electrode composition, Figure 5d displays a main contribution at 529.7 eV reflecting the O 1s in the NCAM lattice, along with further shoulders at 531.4 eV and 532.8 eV typically attributed to the presence of C–O, C=O, and O–H groups deriving from minor contaminations of the sample surface[58,59] and, perhaps, to a defective layered oxide.[60] Moreover, weak Na KLL Auger contributions are observed at 534.2 eV and 541.0 eV.[61] Similarly, the C 1s region (Figure 5e) reveals a major peak at 284.5 eV owing to the C–C bond of the conductive additive and the polymer binder backbone in the composite cathode film, as well as the signals of surface C–O, C=O, and O–C=O contaminants at 286.0 eV, 287.8 eV, and 289.0 eV, respectively.[58,59] As expected, an additional C 1s component due to the C–F bond of PVDF occurs at a BE as high as 290.2 eV.[58,59]

Substantial changes are observed in the XPS response of the cycled cathode (Figure 5f–i). Indeed, Figure 5f reveals an additional signal at about 1075.2 eV most likely due $Na_2O$[62] and a shoulder at 1069.3 eV, which may suggest the presence of $Na_2CO_3$,[63] alongside the main component already described at 1070.7 eV due to NCAM and NaF.[46,57] The F 1s (Figure 5g) and O 1s (Figure 5h) regions evidence further modifications, in line with the Na 1s signal increment and with the above discussed results of elemental quantification. Accordingly, Figure 5g displays a notable intensification of the F 1s contribution ascribed to NaF (684.0 eV) as compared to the F 1s signal of the F–C bond (687.4 eV).[50,51] Moreover, Figure 5h shows: *(i)* an increase in contribution to the O 1s peak of the C–O and C=O components at 531.3 eV,[58,59] which corroborate the hypothesis of possible formation of $Na_2CO_3$ upon cycling; *(ii)* the appearance of an additional shoulder likely due to $Na_2O$ at low BE, i.e., 528.7 eV;[64] and *(iii)* a large raise of the Na KLL Auger signals at 534.7 eV and 540.5 eV[61] reflecting the above mentioned increase in sodium content over the electrode surface (see Table 1). As observed for the pristine cathode, Figure 5h reveals the O 1s in the NCAM lattice at a BE of 530.1 eV, although the related signal is partially masked by those triggered by the various compounds of the EEI.[58,59] The C 1s spectrum of the cycled cathode (Figure 5i) reveals a similar peak arrangement to that of the pristine sample, with C–C, C–O, C=O, O–C=O, and C–F components at 284.5 eV, 286.0 eV, 287.8 eV, 288.7 eV, and 290.2 eV, respectively.[58,59] On the other hand, this latter component attributed to the PVDF binder has a relatively low intensity after cycling (compare the C–F component in Figure 5i with that in Figure 5e), which might suggest that EEI species cover the NCAM oxide and partially mask its XPS signals. Analogously, characteristic signals of NCAM, e.g., those attributed to Mn, Ni and Co, are substantially attenuated after cycling, whilst the Na 2s signal are intensified, as displayed in Figure S8 and discussed in detail in the Supporting Information.

**Figure 5**

In summary, our data suggest that cycling may lead to formation of various insulating precipitates over the layered metal-oxide surface, that is, mostly NaF, $Na_2O$, and $NaCO_3$, hindering the $Na^+$ (de) intercalation in the positive electrode.[28] Notably, these deposits heterogeneously cover the layered cathode particles, so that the initial uniformity in insertion degree across the electrode is gradually lost throughout cycling, thereby inducing cracks and deformations in the NCAM grains. A scheme of this degradation mechanism is shown in Figure 6. Earlier reports have shown that FEC may form a resistive passivation layer incorporating NaF on the cathode surface, which hinders the charge transfer at the interphase thereby adversely affecting the cell capacity. On the other hand, the optimization of the electrolyte formulation may effectively ensure a suitable balance between electrode protection ability and charge transfer kinetics.[50] The cathode morphology certainly plays a key role in promoting parasitic electrolyte-decomposition reactions at moderate and high voltage values.[65,66] In regard to this latter aspect, we remark that the uniform micrometric size of the NCAM particles decreases the electrode surface area, thus increasing the actual current density at the interphase and, therefore, limiting the detrimental oxidation of the electrolyte species.[23] However, the degradation of the electrode/electrolyte interphase observed in this work suggests the need of an enhanced electrolyte formulation with higher anodic stability. Indeed, a careful electrolyte design is essential to ensuring suitable characteristic features for possible use in the battery, *e.g.*, fast ion transport, high thermal stability, and the ability to form a proper interphase on both the anode and the cathode. Among the various formulations proposed so far, solutions of sodium salts such as $NaClO_4$, $NaPF_6$, and NaTFSI in carbonate ester-based solvents have been widely employed with promising results, and various additives have been proposed to enhance the electrode surface chemistry.[67] Despite the beneficial effects of fluoroethylene carbonate as additive stabilizing the anode/electrolyte interphase have been extensively demonstrated, the cathode performance appears to be strongly related to both electrolyte and electrode compositions.[68] In this regard, detrimental side reactions of the electrolyte solution at the interphase with the cathode due to poor electrochemical stability may possibly lead to cell degradation during cycling, as indeed observed in this work. Therefore, various attempts have been made to shift upward the electrolyte decomposition potential thereby enabling efficient $Na^+$ intercalation in high-voltage metal-oxide cathodes. Propylene carbonate (PC), diethyl carbonate (DEC), and ethylene carbonate (EC):PC binary mixtures dissolving $NaClO_4$ typically have satisfactory anodic stability, which can be further improved by FEC addition. Furthermore, earlier studies have suggested that the use $NaPF_6$ and sodium difluoro(oxalato)borate (NaDFOB) may widen the electrochemical stability window (ESW) of the electrolyte solution, leading to decomposition voltage values well above 4.6 V $vs.$ $Na^+$/Na, that is, the voltage cutoff employed in this work (see Figure 1).[69] On the other hand, we point out that a lack of uniformity in the experimental setup for measuring the ESW may hinder a straightforward comparison of the above-mentioned different electrolyte systems.

**Figure 6**

## Conclusions

The relevant changes in microstructure and composition of the NCAM electrode have been thoroughly investigated in this work by adopting a correlative approach to explain the cathode performance in sodium cell upon cycling. X-ray CT data supported by SEM imaging have shown that repeated $Na^+$ intercalation cycles within the layered oxide framework may deform and partially crack in the NCAM phase, thereby leading to a decrease in average particle size. These microstructural modifications have been causally related by XRD analyses to the formation of $Na_{1−x}MO_2$ domains having different insertion degrees, i.e., within a certain range of $x$ values across the cathode. Furthermore, XPS and EDS analyses indicated a predominant presence of NaF, $Na_2O$, and $NaCO_3$ over the cycled electrode, which have been visualized by tomographic reconstructions. These species are likely to have been produced *via* parasitic reactions at the electrode/electrolyte interphase and, in turn, play a key role on battery performance by affecting the local $Na^+$ insertion kinetics, thereby causing partial insulation of the NCAM cathode. According to this model, such a local insulation can cause the observed compositional heterogeneity across the cathode after cycling which induces stresses, distortions of the metal oxide layers, and fractures in the NCAM particles. Notably, the stabilization of the cathode/electrolyte interphase may be crucial for preventing cell degradation processes over cycling, even though irreversible phase transitions in the positive electrode are effectively mitigated by employing optimal stoichiometries. Therefore, enhancing the electrolyte formulation to widen the electrochemical stability window may actually improve the cycle life of the cell. In summary, the results of our study highlight a close interplay between the various sodium battery components in determining a stable electrochemical process and suitable performances.

**Experimental**

NCAM with a mixed P3/P2-type structure, consisting of 79 vol.% P3 (space group $R3m$, No. 16) and 21 vol.% P2 (space group $P6_3/mmc$, No. 194), was synthesized via co-precipitation of hydroxide precursors followed by calcination as previously reported.[23] Electrode disks were prepared *via* doctor blade coating of a dense slurry of NCAM powder dispersed through an agate mortar and a pestle together with polyvinylidene fluoride (Solef® 6020 PVDF), and conductive carbon black (Super P, Timcal) in the ratio 8:1:1 in N-methyl-pyrrolidone (Sigma-Aldrich). The slurry was cast on an aluminum foil (thickness of 15 µm, MTI Corporation) and dried for ca. 3 h on a hot plate. Afterwards, disks with diameter of 14 mm were cut out from the cathode tape, dried overnight under vacuum at 110 °C, and an active material loading of ca. $3.5 \pm 1$ mg cm$^{-2}$ was determined. CR2032 coin-cells (MTI) were assembled by using a sodium-metal disk with diameter of 14 mm, 2 Whatman GF/A glass fiber separators with diameter of 16 mm soaked by the electrolyte solution, and an NCAM electrode disk. The electrolyte solution was prepared in an Ar-filled glovebox (MBraun, $O_2$ and $H_2O$ content below 0.5 ppm) by dissolving 1 M NaClO$_4$ in PC and subsequently adding FEC in a concentration of 20 wt.% with respect to the final solution. The water content in the PC solvent was determined to be lower than 10 ppm by Karl-Fischer titration using an 899 Coulometer (Metrohm). The anode disks were made by rolling and pressing sodium-metal cubes (in mineral oil, 99.9% trace metals basis, Sigma-Adrich) after removing the oil and polishing the metal surface. The cell was charged and discharged over 100 cycles at room temperature (25 °C) using a constant current of 30 mA g$^{-1}$ as referred to the weight of NCAM in the cathode, within the voltage range from 1.4 V to 4.6 V, using a MACCOR series 4000 battery test system. Afterwards, the cell was disassembled in an Ar-filled glovebox (MBraun, $O_2$ and $H_2O$ content below 0.5 ppm), and the cycled cathode was rinsed with anhydrous dimethyl carbonate (DMC) and dried at room temperature under vacuum for few minutes. An additional cathode sample was recovered as above described from a cell charged and discharged over 50 cycles at room temperature using a constant current of 15 mA g$^{-1}$.

SEM, SEM-EDS, XRD, X-ray CT, and XPS analyses were carried out on pristine and cycled electrode samples. SEM-EDS imaging and elemental quantification was performed through a Zeiss EVO MA10 using a tungsten thermionic electron source set at 20 kV and an INCA X-ACT Oxford Instrument analyzer. XRD patterns were collected at a scan rate of 0.5° min$^{-1}$ with a step size of 0.01°, by a Rigaku SmartLab diffractometer using a Cu-Kα source. X-ray CT datasets at the nanoscale were collected by a Zeiss Xradia 810 Ultra instrument (Carl Zeiss Inc.) equipped with a micro-focus rotating Cr anode set at 35 kV and 25 mA (Kα radiation energy of 5.4 keV, MicroMax-007HF, Rigaku). A condenser lens in an elliptical capillary inside a He-filled compartment (condenser chamber) focused the X-ray beam on the sample, while a Fresnel zone plate lens inside an additional He-filled compartment (optics chamber, after transmission through the sample) produced an image of the specimen on a CCD detector. A pinhole positioned at about 7 mm from the specimen mitigated the effects of X-ray scattering from the sample stage. Radiographs of the sample (1901 projections) were taken in absorption-contrast and large-field-of-view (LFOV, 65 μm) mode by rotating the specimen through 180°, with exposure time for each radiograph of 40 s and camera binning 1, which led to a voxel size of 63 nm. Specimens with suitable geometry[70] for X-ray CT at the nanoscale were prepared by laser cutting the electrodes by means of a Series/Compact Laser Micromachining System (Oxford Laser) after peeling off the Al foil with the aid of a razor blade and an optical microscope,[71] and by subsequently gluing these cut portions onto stainless steel (SS) dowels with diameter of 1 mm by epoxy [2,4,6-tris(dimethylaminomethyl)phenol, Devcon]. The electrodes were exposed to air for preparing samples for SEM-EDS and X-ray CT.

XPS data were collected with a Kratos Axis SUPRA that employs monochromated Al Kα (1486.69 eV) X-rays at 15 mA emission and 12 kV HT (180W) and a spot size/analysis area of $700 \times 300$ μm. The instrument was calibrated to gold metal Au 4f (83.95 eV) and dispersion adjusted to give a BE of 932.6 eV for the Cu $2p_{3/2}$ line of metallic copper. Ag $3d_{5/2}$ line full width at half maximum (FWHM) at 10 eV pass energy was 0.544 eV. Source resolution for monochromatic Al Kα X-rays was ~0.3 eV. The instrumental resolution was determined to be 0.29 eV at 10 eV pass energy using the Fermi edge of the valence band for metallic silver, with charge compensation system on <1.33 eV FWHM on PTFE. High resolution spectra were acquired using a pass energy of 20 eV, step size of 0.1 eV and sweep time of 60s, leading to a line width of 0.696 eV for Au $4f_{7/2}$. Survey spectra were obtained with a pass energy of 160 eV. Charge neutralisation was obtained using an electron flood gun with filament current = 0.38 A, charge balance = 2 V, filament bias = 4.2 V. Successful neutralisation was adjudged by analysing the C 1s region wherein a sharp peak with no lower BE structure was obtained. Spectra have been charge corrected to the main line of the carbon 1s spectrum set to 284.8 eV. All data were recorded at a base pressure of below 9 x $10^{-9}$ Torr and a room temperature of 294 K. Data were analysed using CasaXPS v2.3.19PR1.0. Peaks were fitted with a Shirley background prior to analysis.

Tomographic datasets were reconstructed through the Zeiss XMReconstructor software (Carl Zeiss Inc.), which uses a filtered back-projection algorithm. X-ray CT data processing, analysis, and visualization were performed by the Avizo 2020.2 software (Visualization Sciences Group, FEI SAS, Thermo Fisher Scientific). Non-local means and unsharp masking filters were applied to these data, and image segmentation was carried out using grayscale thresholding[24] and watershed[26] methods. Various domains with different density were identified based on differences in attenuation of the X-ray beam: (i) NCAM (high attenuation), (ii) electrode/electrolyte interphase species formed during cycling (indicated by the EEI acronym; moderate attenuation; only detected in the electrode after cycling), (iii) carbon-binder domain (low attenuation), (iv) exterior pores (negligible attenuation as referred to the instrumental reference images). "Continuous PSD" was determined by ImageJ plugin XLib.[24,45] The XPS data were analyzed using the CasaXPS software (Casa Software Ltd).

**Supporting Information**

Supporting Information is available from the Wiley Online Library or from the author.

**Acknowledgements**

This work has received funding from the European Union's Horizon 2020 research and innovation programme Graphene Flagship under grant agreement No 881603, and by the grant "Fondo di Ateneo per la Ricerca Locale (FAR) 2020", University of Ferrara, and performed within the collaboration project "Accordo di Collaborazione Quadro 2015" between University of Ferrara (Department of Chemical and Pharmaceutical Sciences) and Sapienza University of Rome (Department of Chemistry). The X-ray CT measurements were supported by funding from EPSRC (EP/K005030/1). PRS acknowledges the support of The Royal Academy of Engineering (CiET1718/59). PRS and DD acknowledge the support of The Faraday Institution (EP/S003053/1). The XPS data collection was performed at the EPSRC National Facility for XPS ("HarwellXPS"), operated by Cardiff University and UCL, under contract No. PR16195.

**Table captions**

**Table 1.** Elemental composition of cathode samples before (pristine) and after 100 cycles as determined by EDS and XPS.

**Figure captions**

**Figure 1.** Behavior of NCAM over 100 galvanostatic charge/discharge cycles in a sodium cell. In detail: **(a)** trend of specific capacity (left-hand side y axis) and coulombic efficiency (right hand-side y axis); **(b)** voltage profiles and **(c)** differential capacity profiles of the 1st, 50th, and 100th cycles. Electrolyte solution: 1 M NaClO$_4$ in PC, 20 wt.% FEC. Sodium-metal counter electrode. Voltage range: 1.4 − 4.6 V. Current: 30 mA g$^{-1}$ as referred to the active material weight in the cathode. Temperature: 25 °C.

**Figure 2.** **(a, b)** SEM images and **(c–f)** XRD patterns of the NCAM cathode before (pristine) and after cycling in a sodium cell. In detail: **(a, b)** SEM images of **(a)** pristine electrode and **(b)** electrode after 100 cycles; **(c)** XRD patterns in the 2$\theta$ range from 10° to 60° with reference reflections of the P3 layered structure (ICSD # 184736, space group *R3m*, No. 160, † symbol in figure) and P2 layered structure (ICSD # 291156, space group *P6$_3$/mmc*, No. 194, * symbol in figure); **(d–f)** XRD patterns in 2$\theta$ ranges of **(d)** †(003) and *(002) reflections, and of **(e)** †(006) and *(004) reflections, as well as in the 2$\theta$ range **(f)** from 10° to 60°; diffraction patterns collected at a scan rate of 0.5° min$^{-1}$ with step size of 0.01° and using a Cu-K$\alpha$ source. See Figure 1 and the experimental section for further details on the cycling test.

**Figure 3.** X-ray CT imaging at the nanoscale of the NCAM cathode **(a, b, e)** before and **(c, d, f)** after 100 cycles in a sodium cell. In detail: **(a, c)** cross-sectional slices extracted in a plane parallel to the rotation axis (attenuation of X-ray beam depicted by grayscale) and **(b, d)** corresponding segmented slices displaying NCAM (green, highly attenuating), electrode/electrolyte interphase species (EEI, light red, moderately attenuating), carbon-binder domain (gray, lowly attenuating), and exterior/pores (black); **(e, f)** volume rendering of the NCAM phase (green) and of the electrode/electrolyte interphase species (EEI, light red) as reconstructed from the X-ray nano-CT

datasets (carbon and binder in the electrode are not taken into account); panel **f** inset: volume rendering showing individually the NCAM (top image, green) and EEI (bottom image, light red) phases in the cathode. X-ray CT datasets acquired in absorption-contrast and large-field-of-view (LFOV, 65 μm) mode by taking 1901 radiographs of the sample through 180°, with exposure time of 40 s, camera binning 1, and microfocus rotating anode X-ray source set at 35 kV and 25 mA. See Figure 1 and the experimental section for further details on the cycling test.

**Figure 4.** Continuous particle size distribution (PSD) for the NCAM phase in the cathode **(a)** before (pristine) and **(b)** after 100 cycles in a sodium cell, as determined by analyzing X-ray nano-CT datasets.[24,45] See Figures 1 and 3 for further details on the cycling test and on the parameters for X-ray CT acquisition, respectively, as well as the experimental section.

**Figure 5.** XPS of the NCAM cathode before (pristine) and after 100 cycles in a sodium cell. In detail: **(a)** survey spectra with peak indexing and **(b–i)** deconvoluted spectra of **(b, f)** Na 1s, **(c, g)** F 1s, **(d, h)** O 1s, and **(e, i)** C 1s for cathode samples **(b–e)** before (pristine) and **(f–i)** after 100 cycles. See Figure 1 and the experimental section for further details on the cycling test and Figure S8 in the Supporting Information for additional analyses of different regions of the spectra.

**Figure 6.** Scheme of the proposed degradation mechanism of sodium layered oxide cathodes during operation in cell.

| Element | EDS | | | XPS | | |
|---------|-----|-----|-----|-----|-----|-----|
| | Pristine (At.%) | After 100 Cycles (At.%) | $\Delta_{EDS}$% | Pristine (At.%) | After 100 Cycles (At.%) | $\Delta_{XPS}$% |
| O | 40.1±1.9 | 40.7±1.7 | (+2±9)% | 14.14 | 26.37 | +86% |
| C | 27±3 | 3±3 | (−89±13)% | 66.64 | 47.57 | −29% |
| F | 8.4±0.4 | 17.0±1.2 | (+100±20)% | 14.70 | 17.27 | +17% |
| Na | 6.8±0.5 | 16.0±1.0 | (+130±30)% | 1.28 | 8.04 | +529% |
| Ni | 3.9±0.2 | 5.0±0.4 | (+30±20)% | 0.86 | 0.75 | −13% |
| Co | 3.9±0.2 | 5.0±0.4 | (+30±20)% | 0.75 | - | - |
| Mn | 9.1±0.5 | 10.3±0.8 | (+14±15)% | 1.62 | - | - |
| Al | 0.38±0.03 | 0.62±0.08 | (+60±30)% | - | - | - |

**Table 1**

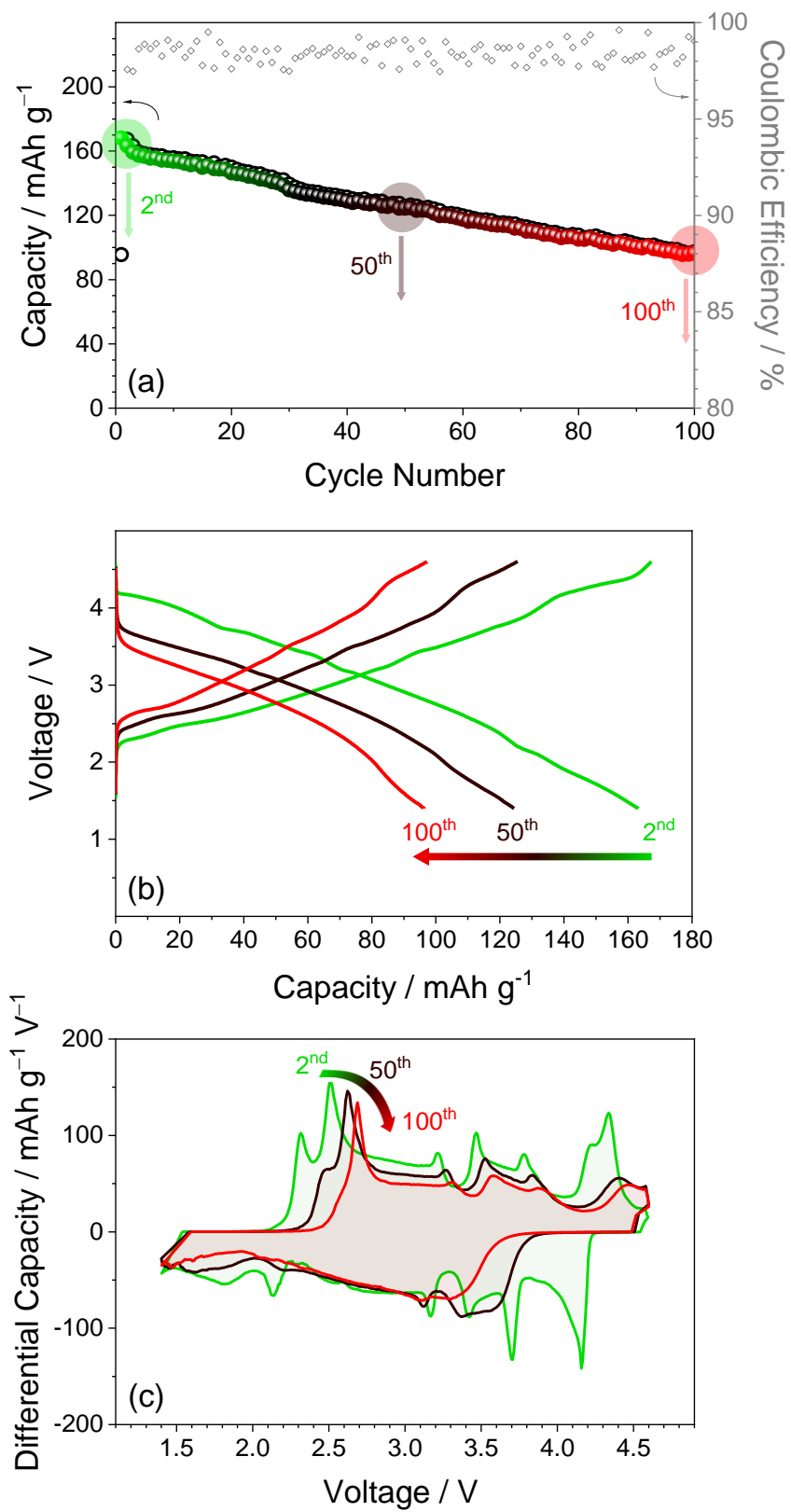

**Figure 1**

**Figure 2**

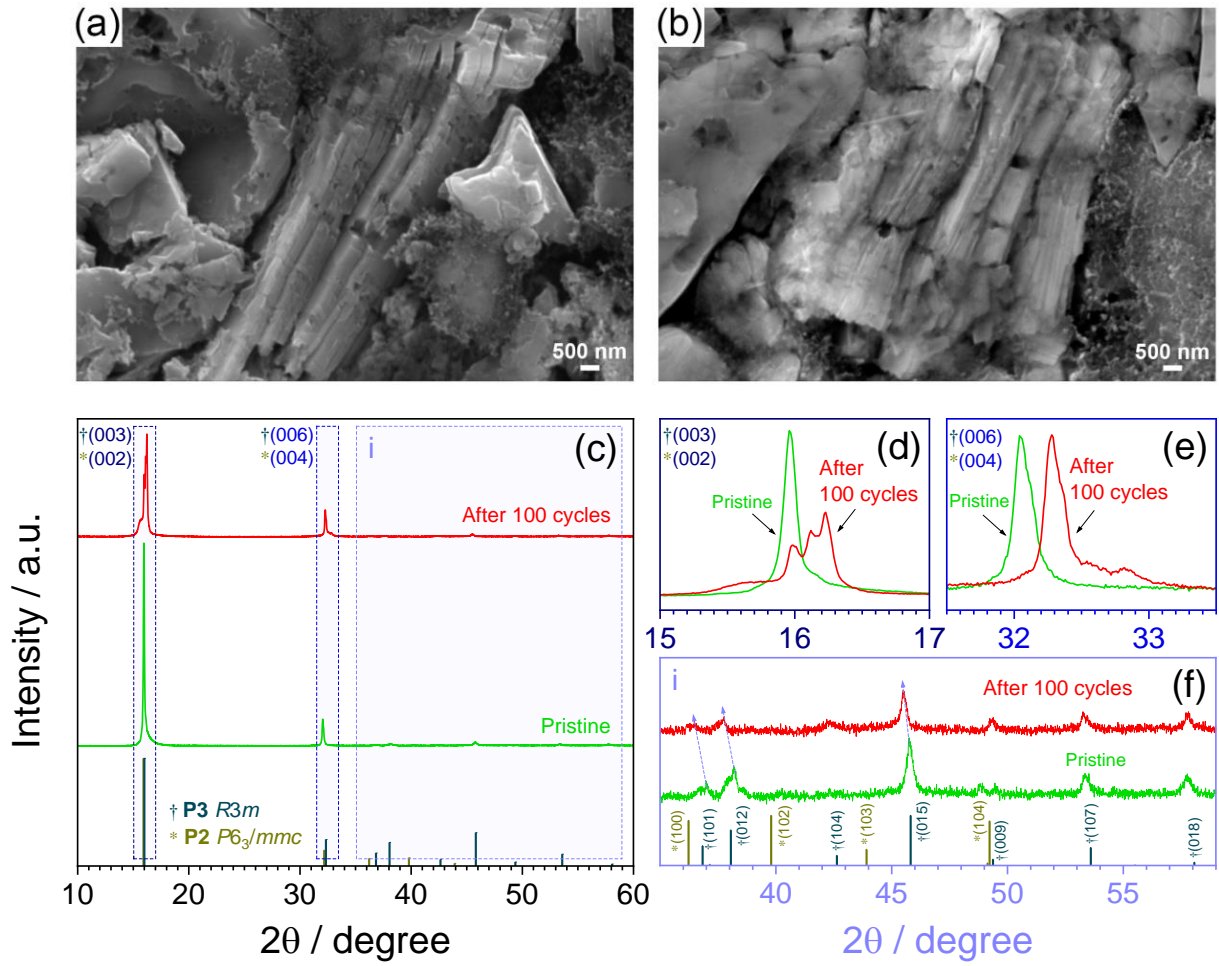

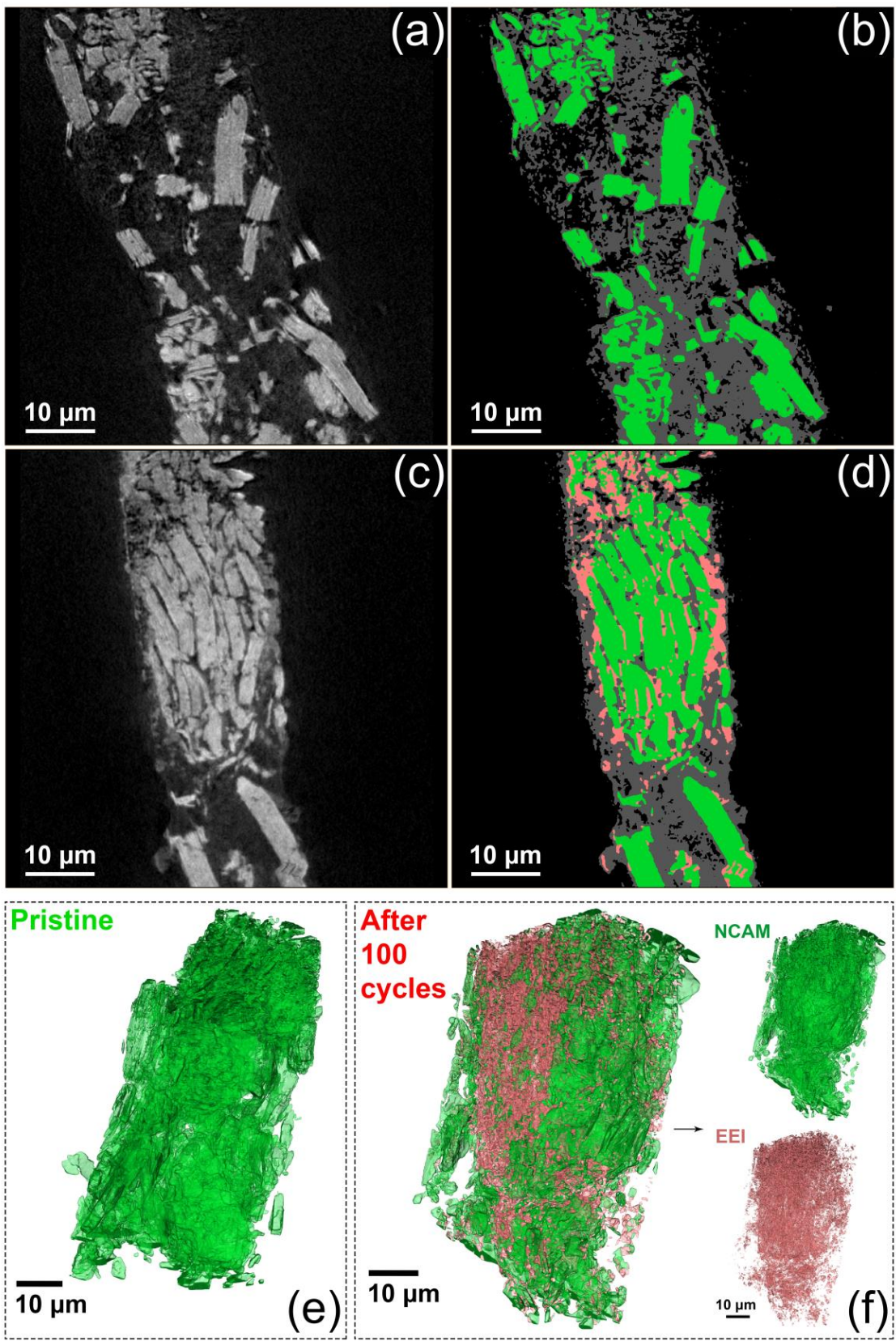

**Figure 3**

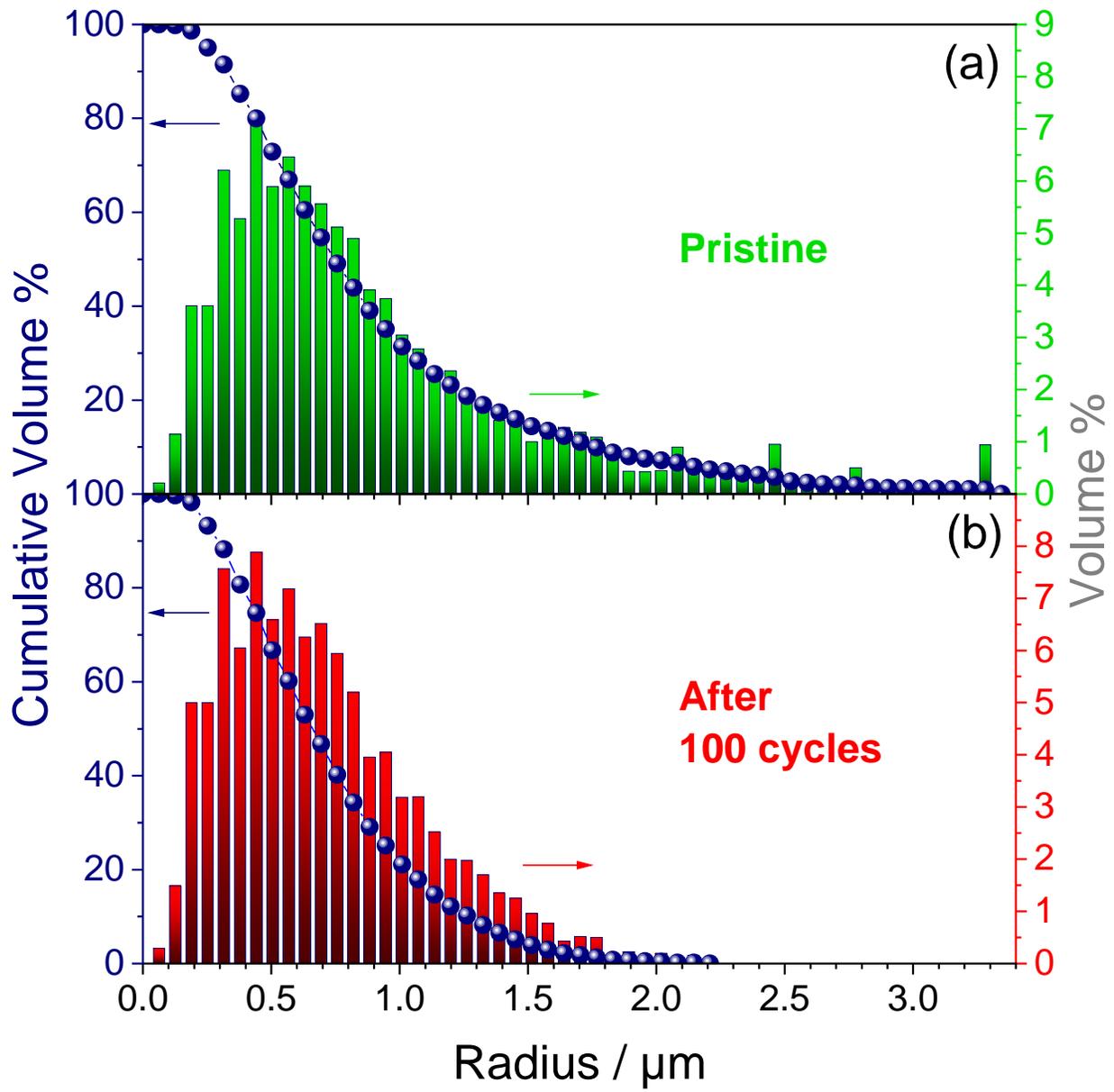

**Figure 4**

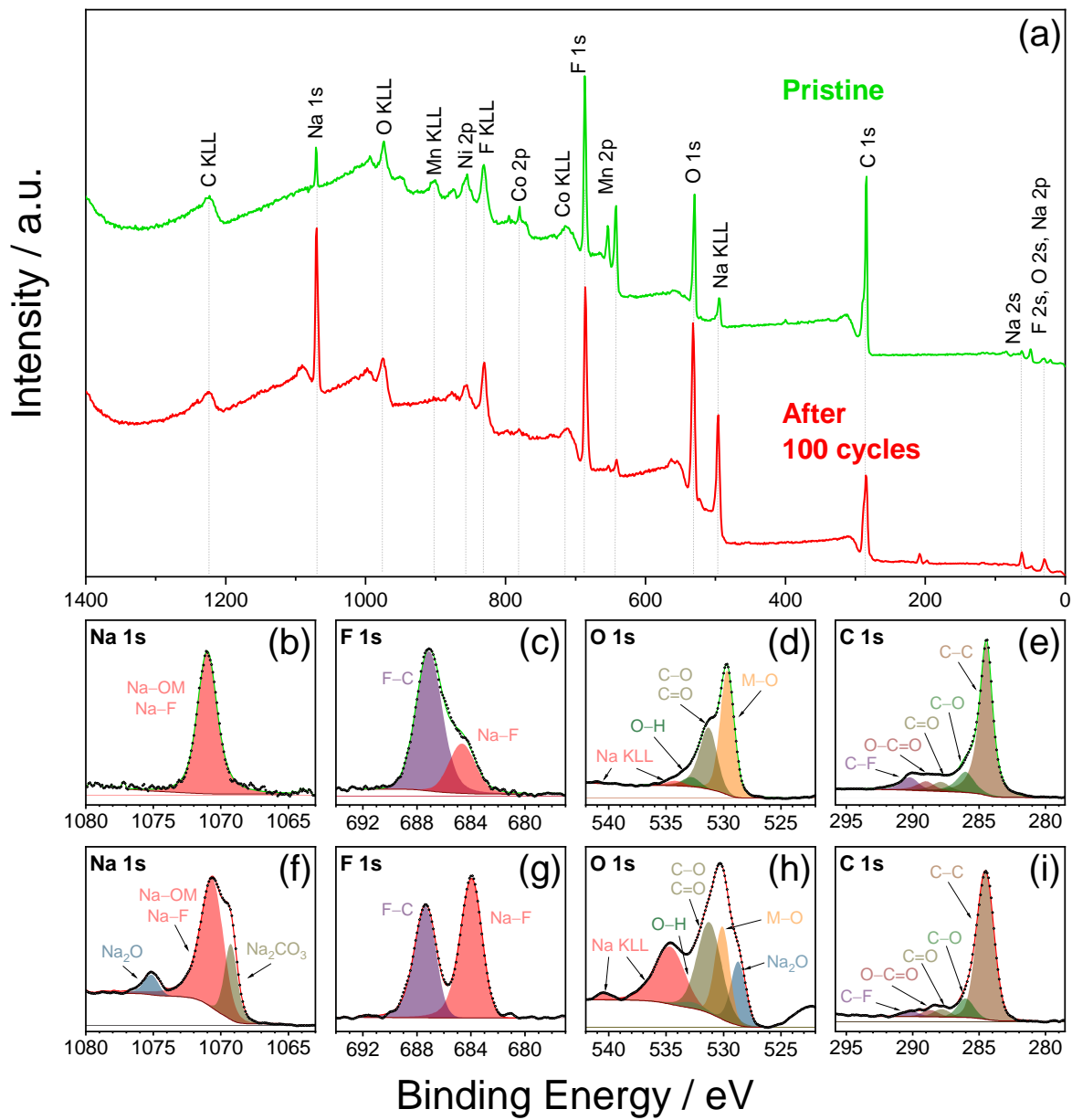

**Figure 5**

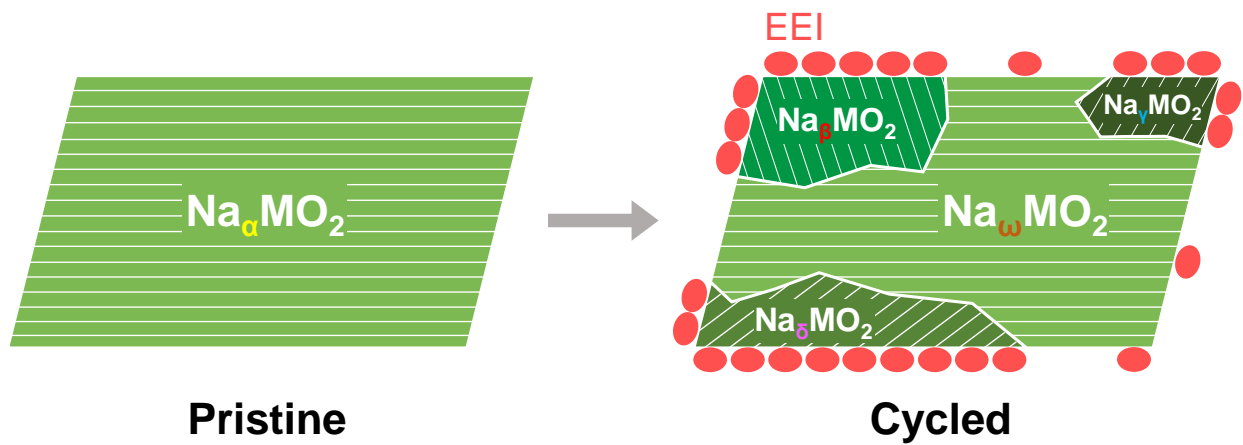

**Figure 6**

# Degradation of layered oxide cathode in a sodium battery: a detailed investigation by X-ray tomography at the nanoscale

Daniele Di Lecce,[a,b,c] Vittorio Marangon,[d] Mark Isaacs,[e,f] Robert Palgrave,[e,f] Paul R. Shearing,[b,c] Jusef Hassoun[a,d,g*]

[a] *Graphene Labs, Istituto Italiano di Tecnologia, via Morego 30, Genova, 16163, Italy*

[b] *Electrochemical Innovation Lab, Department of Chemical Engineering, UCL, London, WC1E 7JE, United Kingdom*

[c] *The Faraday Institution, Quad One, Becquerel Ave, Harwell Campus, Didcot, OX11 0RA United Kingdom*

[d] *University of Ferrara, Department of Chemical and Pharmaceutical Sciences, Via Fossato di Mortara 17, Ferrara, 44121, Italy*

[e] *Department of Chemistry, UCL, 20 Gordon St, Bloomsbury, London, WC1H 0AJ, United Kingdom*

[f] *HarwellXPS, Research Complex at Harwell, Rutherford Appleton Laboratories, Harwell, Didcot, OX11 0FA, United Kingdom*

[g] *National Interuniversity Consortium of Materials Science and Technology (INSTM), University of Ferrara Research Unit, University of Ferrara, Via Fossato di Mortara, 17, 44121, Ferrara, Italy.*

* Corresponding author: jusef.hassoun@unife.it, jusef.hassoun@iit.it.

# Supporting Information

Galvanostatic cycling in the sodium cell leads to significant structural changes in the $Na_{0.48}Al_{0.03}Co_{0.18}Ni_{0.18}Mn_{0.47}O_2$ (NCAM) cathode, as thoroughly discussed in the article (see Figure 2 in the results and discussion section). Electron microscopy of the cathode after 50 cycles provides further insight on the time evolution of the electrode/electrolyte interphase. Figure S1 shows an SEM image of the NCAM cathode, which suggests layered particles surrounded by a carbon-binder composite, in full agreement with the results of Figure 2a and b in the article.

Figure S2a shows the X-ray diffraction (XRD) patterns of pristine and cycled electrode samples in the $2\theta$ region of the (003) and (002) reflections of the P3 and P2 phases,[1] which reflect the $MO_6$ interlayer distance. Multiple peaks occur in the diffractogram of the cycled cathode, that is, at 16.0°, 16.1°, and 16.2°, along with a shoulder at 15.6°. This evidence may indicate either inhomogeneous sodium intercalation degree in the mixed P3/P2 material[1] or irreversible nucleation of new crystal phases upon cycling, such as O3, O1, and P1 domains.[2,3] On the other hand, the XRD patterns between 30° and 60° (Figure S2b) exclude the latter hypothesis, and reveal that the major P3 phase is retained after 100 cycles. It is worth mentioning that traces of P1 domains might nucleate at low sodium intercalation degree by slight distortion of the P3 framework and remain in the cathode after cycling.[3]

Scanning electron microscopy (SEM) images and energy dispersive X-ray spectroscopy (EDS) maps reveal the changes in morphology and elemental composition of the electrode upon cycling (Figure S3). The pristine cathode incorporates the NCAM particles within a carbon-binder matrix, as shown in Figure S3a, and is characterized by a homogenous distribution of the various elements, as suggested by Figure S3b–h. Instead, the electrode after 100 cycles presents additional agglomerates on the surface (see Figure S3i) which have a different composition as compared to the surrounding particles (see Figure S3j–p). These species are further visualized in X-ray computed tomography (CT) reconstructions in the results and discussion section of the article. Qualitative and quantitative analyses reveals that these aggregates mainly consist of NaF, $Na_2O$, and $NaCO_3$ precipitates. These evidences are reported and thoroughly discussed in the article.

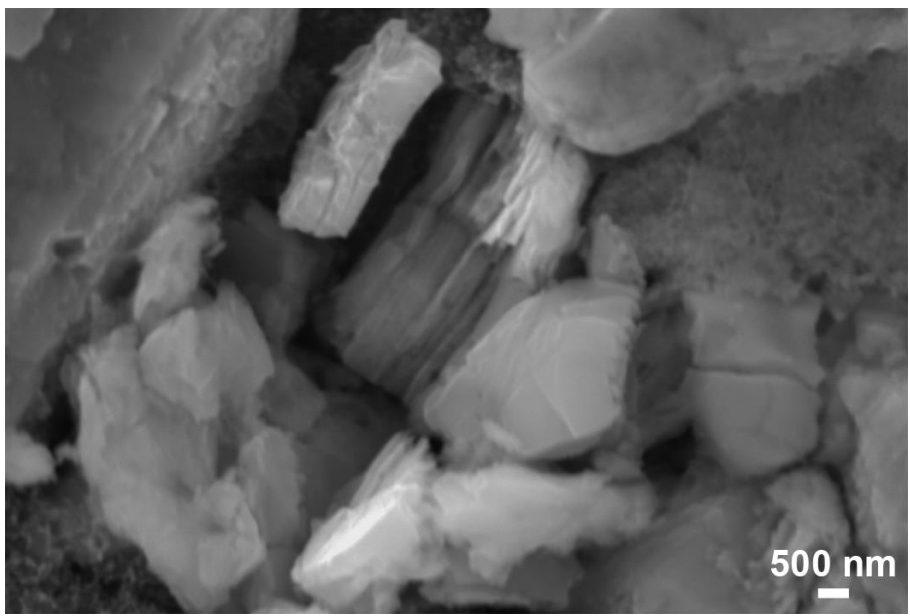

**Figure S1.** SEM image of the NCAM cathode after 50 cycles in a sodium cell. See the experimental section of the article for further details on the cycling test.

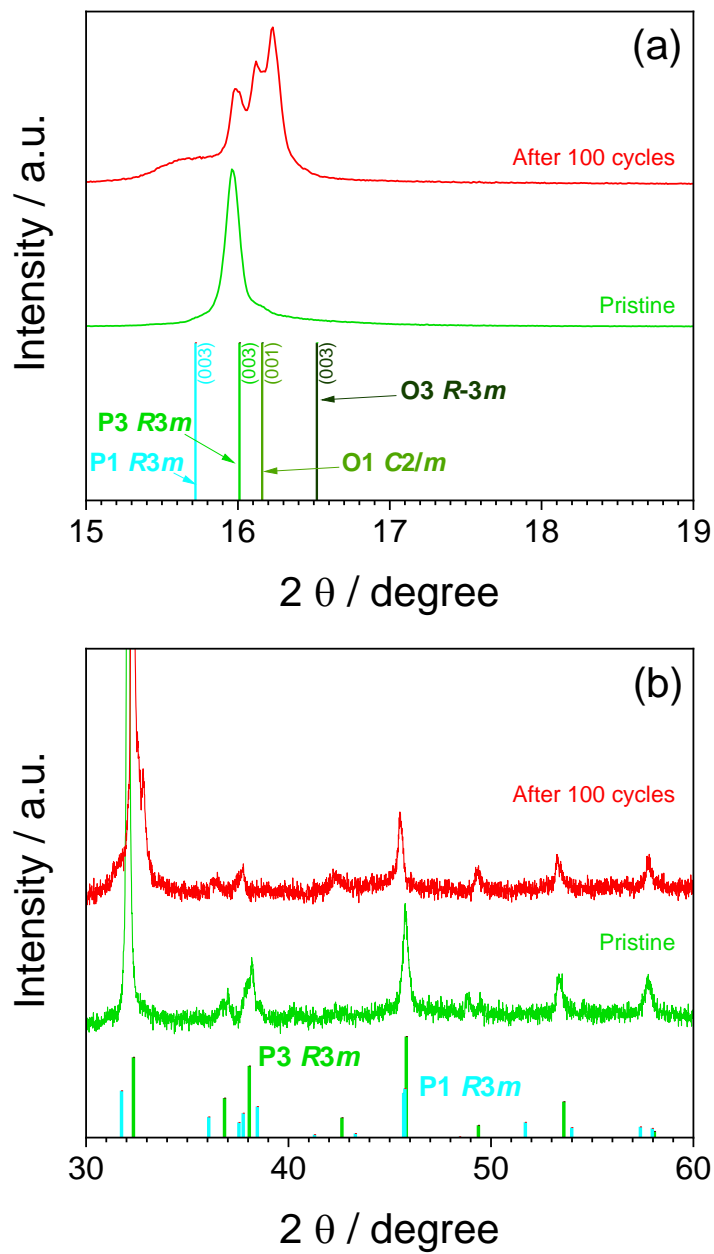

**Figure S2.** XRD patterns of the NCAM cathode before (pristine) and after 100 cycles in a sodium cell, within the 2$\theta$ ranges **(a)** from 15° to 19° and **(b)** from 30° to 60°. The reference reflections of the P3 (ICSD # 184736, space group $R3m$, No. 160), P2 (ICSD # 291156, space group $P6_3/mmc$, No. 194, not reported here for simplicity), P1 (ICSD # 184737, space group $R3m$, No. 160), O3 (ICSD # 184734, space group $R\overline{3}m$, No. 166), and O1 (ICSD # 184735, space group $C2/m$, No. 12) structures are also shown. Diffraction patterns collected at a scan rate of 0.5° min$^{-1}$ with step size of 0.01° and using a Cu-K$\alpha$ source. See Figure 1 in the results and discussion section of the article as well as the related experimental section for further details on the cycling test.

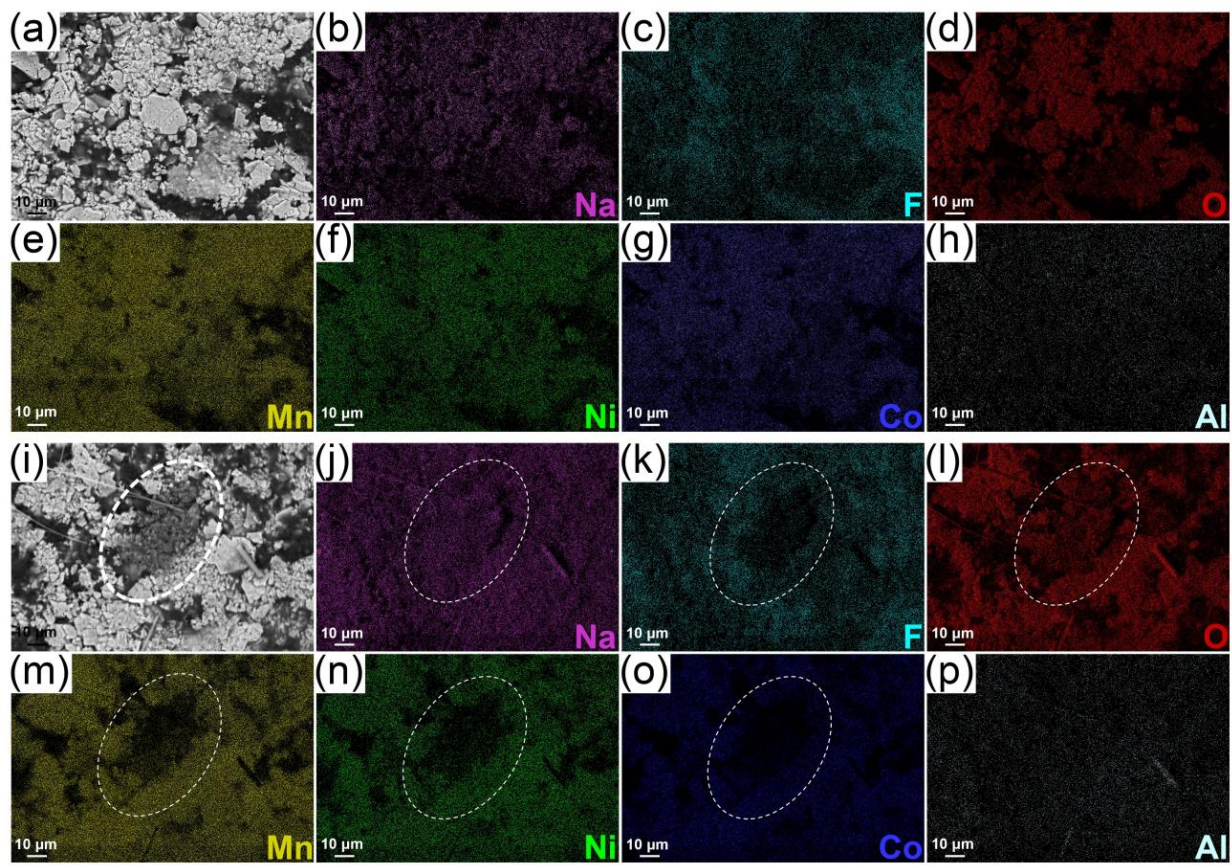

**Figure S3.** **(a, i)** SEM images in the backscattered electron mode and corresponding **(b–h, j–p)** EDS maps of the NCAM cathode **(a–h)** before (pristine) and **(i–p)** after 100 cycles in a sodium cell. Elemental maps showing the distribution of **(b, j)** Na, **(c, k)** F, **(d, l)** O, **(e, m)** Mn, **(f, n)** Ni, **(g, o)** Co, and **(h, p)** Al. An additional agglomerate formed on the surface of the cycled electrode is marked in panels **i–p**. See Figure 1 in the results and discussion section of the article as well as the related experimental section for further details on the cycling test.

The NCAM cathode has been reconstructed at the nanoscale before and after cycling by X-ray CT. The local attenuation of the X-ray beam by the specimen enables us to identify the various electrode components based on the density of the different phases and to perform quantitative analyses on their spatial distribution across the field of view.[4] These data are reported in Figures 3 and 4 in the article and in Supporting Movies S1–S4. The curves represented in Figure

S4 show the relative attenuation of the NCAM phase, the electrode/electrolyte interphase species (EEI), the carbon-binder domain (CBD), and the exterior/pores.

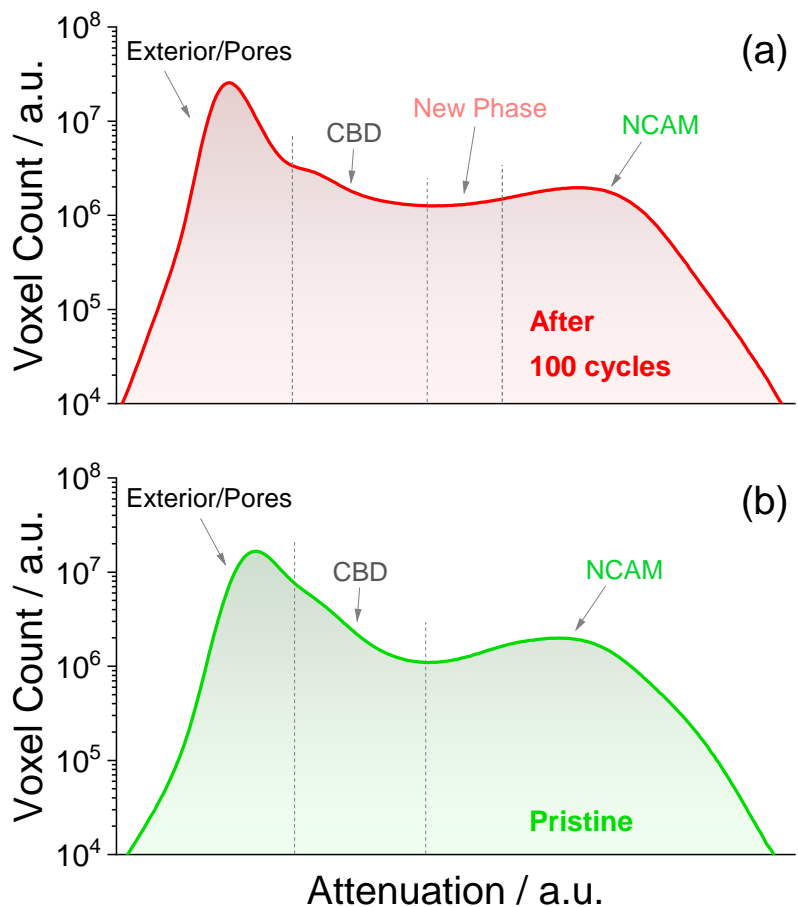

**Figure S4.** Distribution curve of the local attenuation of the X-ray CT beam by the electrode specimens **(a)** after 100 cycles and **(b)** in pristine condition. X-ray CT datasets acquired in absorption-contrast and large-field-of-view (LFOV, 65 µm) mode by taking 1901 radiographs of the sample through 180°, with exposure time of 40 s, camera binning 1, and microfocus rotating anode X-ray source set at 35 kV and 25 mA. See Figure 1 in the results and discussion section of the article as well as the related experimental section for further details on the cycling test conditions. See additional X-ray CT results in Figures 3 and 4 in the article, Figures S5–S7, and Supporting Movies S1–S4 along with the related discussion.

X-ray CT, SEM-EDS, and XPS analyses reveal the precipitation of inorganic compounds at the EEI, as discussed above as well as in the article. Figure S5 reveals that the layered metal-oxide particles after 50 cycles have heterogeneous size and are surrounded by the EEI domain, as already described in the article for the electrode after 100 cycles.

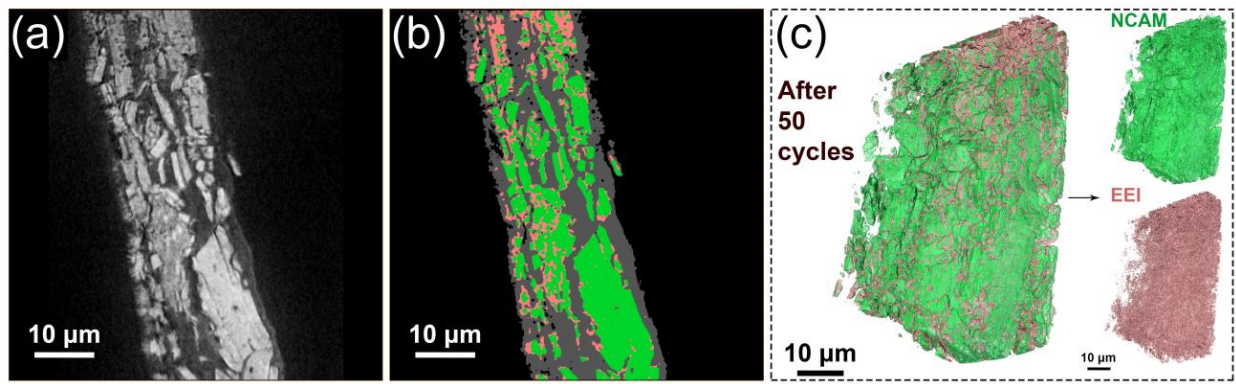

**Figure S5.** X-ray CT imaging at the nanoscale of the NCAM cathode after 50 cycles in a sodium cell. In detail: **(a)** cross-sectional slice extracted in a plane parallel to the rotation axis (attenuation of X-ray beam depicted by grayscale) and **(b)** corresponding segmented slices displaying NCAM (green, highly attenuating), electrode/electrolyte interphase species (EEI, light red, moderately attenuating), carbon-binder domain (gray, lowly attenuating), and exterior/pores (black); **(c)** volume rendering of the NCAM phase (green) and of the electrode/electrolyte interphase species (EEI, light red) as reconstructed from the X-ray nano-CT datasets (carbon and binder in the electrode are not taken into account); panel **c** inset: volume rendering showing individually the NCAM (top image, green) and EEI (bottom image, light red) phases in the cathode. X-ray CT datasets acquired in absorption-contrast and LFOV (65 µm) mode by taking 1901 radiographs of the sample through 180°, with exposure time of 40 s, camera binning 1, and microfocus rotating anode X-ray source set at 35 kV and 25 mA. See the experimental section of the article for further details on the cycling test conditions.

As discussed in the article, X-ray CT indicates a gradual electrode degradation during cycling, which is reflected as growth of EEI species along with decrease in particle size for the NCAM phase. We have herein further supported our conclusions by analyzing the "continuous particle size distribution" (PSD) of the NCAM phase after 50 cycles (Figure S6). In full agreement with the data of Figure 4 in the article, the results of this analysis confirm that cycling leads to a gradual decrease in particle size.

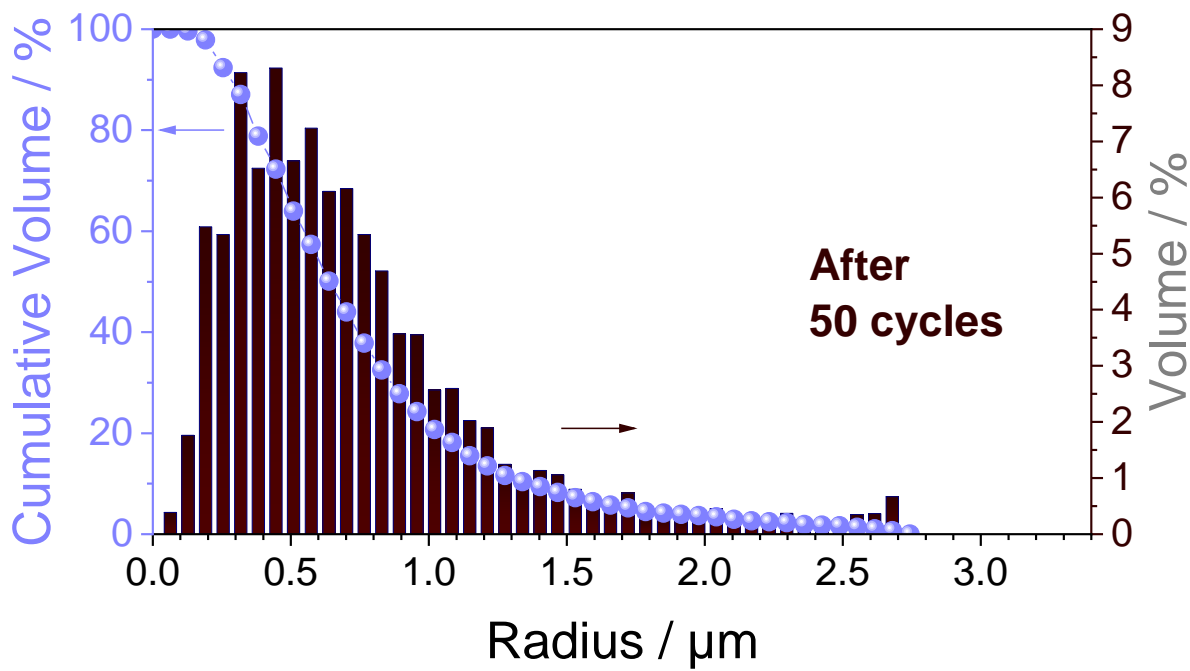

**Figure S6.** Continuous PSD for the NCAM phase in the cathode after 50 cycles in a sodium cell, as determined by analyzing X-ray nano-CT datasets.[4,5] See the experimental section of the article and Figure S5 for further details on the cycling test and on the parameters for X-ray CT acquisition.

The tomographic reconstruction of the electrode after 100 cycles shows that about 15 vol.% of EEI particles in the cathode sample are smaller than 400 nm, as determined by the method proposed by Münch et al.[4,5] and shown in Figure S7.

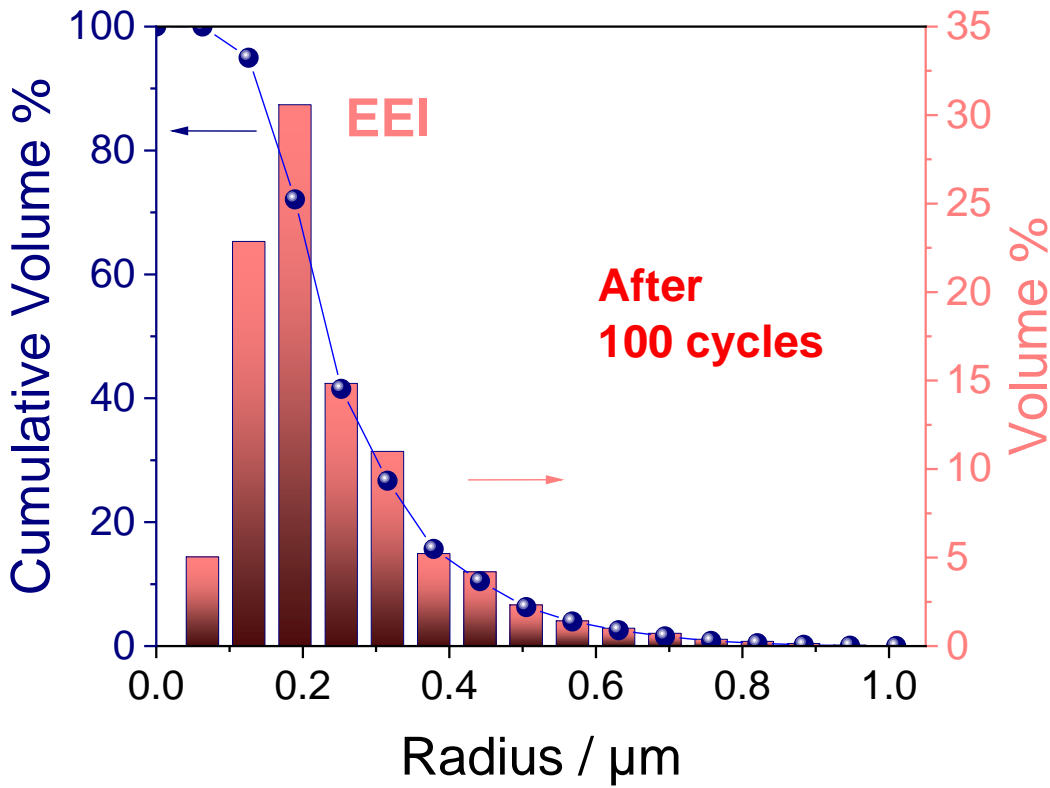

**Figure S7.** Continuous PSD for the EEI species in the cathode after 100 cycles in a sodium cell, as determined by analyzing X-ray nano-CT datasets.[4,5] See Figure 1 in the results and discussion section of the article as well as the related experimental section for further details on the cycling test. See additional X-ray CT results in Figures 3 and 4 in the article, Figures S4–S6, and Supporting Movies S1–S4 along with the related discussion.

XPS suggests that the electrode is partially covered by inorganic compounds containing Na, F, and O, which attenuate the characteristic signals of NCAM. In this regard, the analysis of the survey spectra of Figure 5a and of the selected regions of the Na 1s, F 1s, O 1s, and C 1s photoelectron signals shown in the deconvoluted spectra in Figure 5b–i (see the results and discussion section of the article) provides qualitative and quantitative information on the EEI nature. Besides, the regions displayed in Figure S8 reveal that the signals of Mn, Ni and Co are significantly masked by the new species deposited on the electrode. On the other hand, the figure shows that the Na 2s peak raises significantly after cycling, according to the deposition of NaF, Na$_2$O, and NaCO$_3$ discussed above as well as in the article.

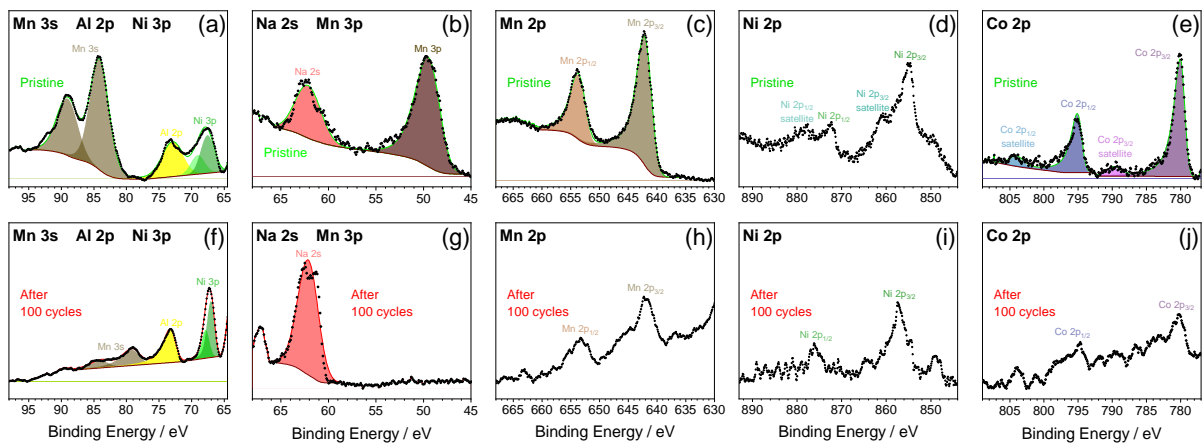

**Figure S8.** Deconvoluted X-ray photoelectron spectra of **(a, f)** Mn 3s, Al 2p, Ni 3p, **(b, g)** Na 2s, Mn 3p **(c, h)** Mn 2p, **(d, i)** Ni 2p, and **(e,j)** Co 2p for cathode samples **(a–e)** before (pristine) and **(f–j)** after 100 cycles in a sodium cell. See Figure 5 and Table 1 in the article for the relevant survey spectra and for additional analyses of different regions of the spectra, respectively, as well as Figure 1 in the article and the related experimental section for further details on the cycling test conditions.